\input harvmac
\input epsf.sty

\Title{\vbox{\baselineskip12pt\hbox{UICHEP-TH/96-04}}}
{\vbox{\centerline{Simple Baryon-Meson Mass Relations} 
\vskip 6pt
\centerline{From A Logarithmic Interquark Potential}}}

\centerline{Tom D. Imbo\footnote{$^*$}{imbo@uic.edu}}

\bigskip\centerline{Department of Physics}
\centerline{University of Illinois at Chicago}
\centerline{845 W. Taylor St.}
\centerline{Chicago, IL \ 60607-7059}

\vskip 0.75in

I consider the quantity $\delta(m_1m_2m_3)\equiv M_{q_1q_2q_3}-(M_{q_1{\bar q}
_2}+M_{q_2{\bar q}_3}+M_{q_1{\bar q}_3})/2$, where the $M$'s represent the 
ground state spin-averaged hadron masses with the indicated quark content and 
the $m$'s the corresponding constituent quark masses. I assume a logarithmic 
interquark potential, the validity of a nonrelativistic approach, and various 
standard potential model inputs. Simple scaling arguments then imply that the 
quantity $R(x)\equiv\delta(mmm_3)/\delta (m_0m_0m_0)$ depends only on the 
ratio $x=m/m_3$, and is independent of $m_0$ as well as any parameters 
appearing in the potential. A simple and accurate analytic determination of 
$\delta (mmm_3)$, and hence $R(x)$, is given using the $1/D$ expansion where 
$D$ is the number of spatial dimensions. When applicable, this estimate of 
$R(x)$ compares very well to experiment --- even for hadrons containing light 
quarks. A prediction of the above result which is likely to be tested in 
the near future is $M_{\Sigma_b^*}/2+(M_{\Lambda_b}+M_{\Sigma_b})/4=5774\pm 4
\ {\rm MeV/c^2}$.

\Date{}

In 1983, Nussinov \ref\nuss{S.~Nussinov, Phys. Rev. Lett. {\bf 51} 
(1983) 2081.} suggested that 
\eqn\one{M_{q_1q_2q_3}>(M_{q_1{\bar q_2}}+M_{q_2{\bar q_3}}
+M_{q_1{\bar q_3}})/2} 
is a rigorous inequality in QCD. Here $M_{q_1q_2q_3}$ represents the mass of 
any spin state of the ground state baryon with the indicated quark content, 
and $M_{q_i{\bar q}_j}$ the mass of the ground state meson in the spin state 
imposed by the chosen baryon.\foot{This inequality was also discussed 
independently in the context of nonrelativistic potential models by Richard 
\ref\rich{J.-M.~Richard, Phys. Lett. B {\bf 139} (1984) 408.}. The special 
case of equal quark masses was first treated by Ader, Richard and Taxil
\ref\ader{J.-P.~Ader, J.-M.~Richard and P.~Taxil, Phys. Rev. D {\bf 25} 
(1982) 2370.}. Similar inequalities were also derived at this time by 
Weingarten \ref\wein{D.~Weingarten, Phys. Rev. Lett. {\bf 51} (1983) 1830.} 
and Witten \ref\witt{E.~Witten, Phys. Rev. Lett. {\bf 51} (1983) 2351.}. 
Various ``improved'' versions of \one\ have also been discussed over the 
years. See, for example, \ref\richt{J.~M.~Richard, Phys. Rep. {\bf 211} 
(1991) 1.} and references therein.} He was able to demonstrate this assuming 
a separation of the three-body Hamiltonian for the quarks in the baryon into 
a sum of two-body pieces --- a result which holds at both weak and strong 
coupling in QCD. Although this does not constitute a complete proof, the 
inequality \one\ is indeed satisfied by the observed ground state baryons and 
mesons. The arguments in \nuss\ apply not only to specific spin states as 
noted above, but also to spin-averaged states \rich . That is, we can take 
the $M$'s in \one\ to represent the spin-averaged ground state baryon and 
meson masses. This view of the Nussinov inequality, which we adopt below, is 
also supported by experiment. 

Is it possible to further understanding the relationship between 
ground state baryon and meson masses? More precisely, if we write 
\eqn\two{M_{q_1q_2q_3}=(M_{q_1{\bar q_2}}+M_{q_2{\bar q_3}}
+M_{q_1{\bar q_3}})/2 +\delta(m_1m_2m_3),}
can we say anything about the positive quantity $\delta(m_1m_2m_3)$? How does 
$\delta$ depend on the quark masses $m_1$, $m_2$ and $m_3$? Does anything 
special happen when one or more of these masses get very large, or when one or 
more are equal? To begin to answer these questions clearly requires more 
specific dynamical input than that used in arguing for the inequality \one .
A first principles analytic computation from QCD is clearly beyond our current
theoretical resources. Numerical results from lattice QCD have not yet reached
the level of accuracy required to address these questions seriously. So, in 
order to proceed at all, a more phenomenological approach is called for. For 
instance, we can attempt to study $\delta$ in the context of potential models.
That is, we can postulate a specific form for the interaction between the quarks
in mesons and baryons, and then solve the appropriate two and three body 
problems to obtain the desired spectrum. Of course, there are necessarily free 
parameters in this approach which must be fit to a subset of the experimental 
data. This is because we do not know the exact form of the interquark 
potential, nor do we know the exact constituent quark masses that should appear
in the above Hamiltonians. Moreover, three body equations are notoriously 
difficult to solve or approximate analytically, so that one must in general 
resort to sophisticated numerical methods. So even this very natural idea of 
potential models, although having been applied very successfully for over 30 
years, is found to lack a certain simplicity and elegance.

We will find, however, that a simple, accurate, and completely analytic
approximation to $\delta$ can be obtained if one assumes a logarithmic
interquark potential and nonrelativistic dynamics governed by the Schr\"odinger
equation --- at least when two of the quark masses are equal. The log 
potential approach to the hadron spectrum has been around for two
decades \ref\quig{C.~Quigg and J.~Rosner, Phys. Lett. B {\bf 71} (1977) 153; 
Phys. Rep. {\bf 56} (1979) 167.}. It was originally motivated by the 
observation that the $\psi (2S)$-$\psi (1S)$ and $\Upsilon (2S)$-$\Upsilon 
(1S)$ mass splittings are approximately equal, as this splitting is quark mass 
independent in the log potential. After many years, and the discovery of many 
new states, fits to the mass spectrum of mesons containing $b$, $c$ and $s$
quarks still support the quasi-logarithmic nature of the interquark potential 
\ref\mart{A.~Martin, Phys. Lett. B {\bf 100} (1981) 511\semi A.~K.~Grant, 
J.~L.~Rosner and E.~Rynes, Phys. Rev. D {\bf 47} (1993) 1981.}. There is also 
evidence supporting this picture in the low-lying baryon spectrum, although 
computations are more difficult here and experimental results more sparse 
\richt\ref\mb{J.-M.~Richard, Phys. Lett. B {\bf 100} (1981) 515.}. Of course, 
it is clear from QCD that this description must eventually break down at very 
short and very large distance scales. However, more realistic QCD-inspired 
potentials --- such as the famous ``Coulomb-plus-linear'' potential \ref\eich
{E.~Eichten, K.~Gottfried, T.~Kinoshita, K.~D.~Lane and T.-M.~Yan, Phys. Rev.
D {\bf 17} (1978) 3090; ibid. {\bf 21} (1980) 203.} --- are all 
quasi-logarithmic over the distance scales relevant for the low-lying spectrum 
of the observed heavy hadrons \ref\buch{W.~Buchmuller and S.-H.~H.~Tye, Phys. 
Rev. D {\bf 24} (1981) 132.}.\foot{Other physical quantities such as production 
and decay rates, which are sensitive to the wave function at the origin, depend 
on physics at shorter distance scales. Here there are substantial 
differences between the predictions of quasi-logarithmic potentials and more 
realistic ones, even though they have the same ``shape'' near the RMS radii of 
the states. See, for example, \ref\cikm{E.~J.~Eichten and C.~Quigg, Phys. Rev. 
D {\bf 52} (1995) 1726\semi S.~J.~Collins, T.~D.~Imbo, B.~A.~King and 
E.~C.~Martell, hep-ph/9610547 (to appear in Physics Letters B).}.} In short, 
the log potential and its relatives have been very successful.

The approach used here is to approximate $\delta$ to leading order in the $1/D$
expansion, where $D$ is the number of spatial dimensions. That is, the relevant
two and three body Hamiltonians are first generalized from 3 to $D$ spatial 
dimensions, and then we assume that $D$ is large. It is well known that a 
systematic $1/D$ expansion can be developed \ref\mlod{L.~Mlodinow and
N.~Papanicolaou, Ann Phys. (N.Y.) {\bf 128} (1980) 314; {\bf 131} (1981) 1.}
\ref\chat{A.~Chatterjee, Phys. Rep. {\bf 186} (1990) 249.}. There is also 
strong evidence that the leading order term is already very accurate for 
quasi-logarithmic potentials, in both the two and three body cases, even for 
$D=3$. The two body results in the $1/D$ expansion also have the nice property 
that they are completely analytic for all power-law potentials $V(r)=Ar^b$. 
Since $\ell n(r)={\rm lim}_{b\to 0}(r^b-1)/b$, the results are also analytic 
for the log potential. In the three body case this property no longer holds 
for general power-law potentials except in the case of three equal masses. For 
instance, when only two masses are equal, the leading order large-$D$ result 
requires the numerical solution of a transcendental algebraic equation 
\ref\vand{P.~du~T.~van~der~Merwe, Phys. Rev. D {\bf 33} (1986) 3383.}. 
However, for the log limit I will show that a completely analytic solution can 
be found. Moreover, when put together with the two body result, this leads to 
an extremely simple and accurate form for $\delta$. (For three unequal masses 
the situation is more complicated \vand , and will not be treated here. Related 
applications of large-$D$ results to the ground state meson and baryon spectra
can be found in \vand\ref\tomm{P.~du~T.~van~der~Merwe, Phys. Rev. D {\bf 30}, 
(1984) 1596\semi T.~D.~Imbo, Phys. Rev. D {\bf 36} (1987) 3438\semi 
A.~Gonzalez, Few Body Sys. {\bf 13} (1992) 105.}.)

Lets begin with the specific form of the interquark forces. I will assume that 
the potential between a quark and an antiquark in a color singlet meson is 
$V_{q_1{\bar q}_2}(r)=A\ell n(r/r_o)$, where $r$ is the interquark separation 
and $A,r_0>0$ are parameters. I make the further standard assumption 
\richt\ that the quark-quark potential in a baryon (that is, two quarks in an 
overall color antitriplet) is 1/2 of the color singlet quark-antiquark potential 
at equal separation: $V_{q_1q_2}(r)=V_{q_1{\bar q}_2}(r)/2$. (I ignore the 
effects of the direct three body interactions that arise in QCD, as these have 
been shown on many occasions to be negligible for the questions that we are 
considering. See, for example, \richt\ref\tax{J.~M.~Richard and P.~Taxil, Ann. 
Phys. {\bf 150} (1983) 267; Phys. Lett. B {\bf 128} (1983) 453.}.) I 
also assume that the dynamics are governed by the appropriate nonrelativistic 
Schr\"odinger equation. If $H_2\psi=E_{m_im_j}\psi$ and $H_3\phi=E_{m_1m_2m_3}
\phi$ are the two and three body Schr\"odinger equations (for the ground state) 
respectively, then we have $\delta(m_1m_2m_3)=E_{m_1m_2m_3}-(E_{m_1m_2}+
E_{m_2m_3}+E_{m_1m_3})/2$. Throughout, the effects of spin are neglected and 
the results are interpreted as representing the masses and binding energies of 
spin-averaged meson and baryon states. The simple rescaling $r_i\to\hbar r_i/
\sqrt{m_iA}$ of the quark position vectors in $H_2$ and $H_3$ shows that
\eqn\three{\eqalign{E_{m_im_j}&=(A/2)[\ell n(\hbar^2/\mu Ar_0^2)+C]\cr 
E_{m_1m_2m_3}&=(3A/4)[\ell n(\hbar^2/m_1Ar_0^2)+f(w,x)],}}
where $w=m_1/m_2$, $x=m_1/m_3$, and $\mu =m_im_j/(m_i+m_j)$ is the two body 
reduced mass. Both the constant $C$ and the function $f$ are independent of 
$\hbar$, $A$ and $r_0$. These scaling relations further imply that
\eqn\four{\delta (m_1m_2m_3)=Ag(w,x),}
where again $g$ is independent of {\it any} dimensionful parameter in the 
problem other than the constituent quark masses, and even this dependence is 
only through the two dimensionless ratios $w$ and $x$. 

Note that an immediate consequence of \four\ is that in the case of equal
quark masses $m_1=m_2=m_3\equiv m$, $\delta$ is independent of $m$. Assuming
that the constituent masses of the $u$ and $d$ quarks are identical and equal
to some value $m_n$, where here and below the symbol $n$ denotes either $u$
or $d$, we may then determine $\delta (m_nm_nm_n)$ from experiment. More 
precisely, $\delta (m_nm_nm_n)=M_{nnn}-3(M_{n{\bar n}})/2$. Assuming further
the standard technique for constructing the spin-averaged masses, we have 
$M_{nnn}=(M_N+M_{\Delta})/2$ and $M_{n{\bar n}}=(M_{\pi}+3M_{\rho})/4$. 
Putting this all together with the measured masses\foot{Unless otherwise 
stated, all experimental results are taken from \ref\mont{Particle Data Group, 
Phys. Rev. D {\bf 54} (1996) 1.}. I also average over the electromagnetic 
splittings within a given isospin multiplet.} of the $N$, $\Delta$, $\pi$ and 
$\rho$ states gives $\delta (m_nm_nm_n)=171\pm 2$ ${\rm MeV/c^2}$. Thus, we 
should expect $\delta (m_sm_sm_s)$ to also be in this range, where $m_s$ is 
the mass of the strange quark. Unfortunately, this cannot be tested directly 
since the spin 1/2 ground state baryon with three $s$ quarks does not exist. 
There is only the spin 3/2 $\Omega^-$. (This is because the spin wave function 
of the identical $s$ quarks must be totally symmetric in the ground state.) 
There is a similar problem in the meson sector where only the spin 1 $\phi$ 
can be interpreted as a nearly pure $s{\bar s}$ ground state. (The spin 0 
combination mixes strongly with $u{\bar u}$ and $d{\bar d}$ states.) Thus, 
some additional theoretical input is needed to obtain the spin-averaged baryon 
and meson masses. However, any reasonable model of the spin-spin interactions 
between quarks, when applied to the $\Omega^-$ and $\phi$ in order to 
reconstruct the spin-averaged states, will yield a value of $\delta (m_sm_sm_s)$ 
around the desired range. A similar test of $\delta (mmm)$ for charm and 
bottom quarks cannot be made since the heavy $ccc$ and $bbb$ baryons have not 
yet been observed.

It is interesting to note that a numerical evaluation of $\delta$ in the case
of equal quark masses yields $\delta (mmm)\simeq (0.225)A$ \mb . (That is, 
$g(1,1)\simeq 0.225$.) If we then use this result to ``fit'' $A$ to the 
observed $\delta$ for $u$ and $d$ quarks, we obtain $A=760\pm 10$ MeV. This is 
not too far from the value $A=733$ MeV \quig\ obtained from a fit of the log 
potential spectrum to meson states containing only heavy quarks! Not only do we 
see a striking consistency between the meson and baryon sectors, but we also 
gain some faith in the application of our nonrelativistic model to hadrons 
containing ``light'' quarks. This should not be too surprising since 
nonrelativistic models have often been successful on such semirelativistic 
quark systems in the past. The standard philosophy regarding this is that 
many relativistic corrections can be incorporated into a nonrelativistic model 
by a redefinition of the constituent quark masses and simple modifications of 
the potential such as the addition of an overall constant, at least for 
low-lying states \ref\basd{J.~Basdevant and S.~Boukraa, Z. Phys. C {\bf 28}
(1985) 413.}. (Note that both of these modifications do not affect the 
result for $\delta (mmm)$ in the log potential.) Moreover, in our case there 
may be some cancellations of additional relativistic corrections occurring 
between the two body and three body contributions to $\delta$. I will maintain 
this viewpoint in what follows and continue to apply the results of the above 
model to systems containing light and/or heavy quarks.

What about the case when only two of the quark masses are equal? Here, 
$\delta (mmm_3)=Ag(1,x)$, where $x=m/m_3$. Instead of resorting to a 
time-consuming numerical determination of $g$ as a function of $x$, lets 
consider $\delta$ to leading order in the $1/D$ expansion. To this end, 
let $H^{(D)}_2\psi=E^{(D)}_{m_im_j}\psi$ and $H^{(D)}_3\phi=
E^{(D)}_{m_1m_2m_3}\phi$ be the generalizations of the two and three body 
Schr\"odinger equations to an arbitrary number of spatial dimensions. That is, 
we keep the interquark potentials the same and simply replace each 
three-dimensional Laplacian appearing in the kinetic energy terms by its 
$D$-dimensional counterpart. We can still write $E^{(D)}_{m_im_j}$ and 
$E^{(D)}_{m_1m_2m_3}$ as in \three , only now the constant $C$ and the 
function $f(w,x)$ depend on $D$. These binding energies can be obtained in 
the leading order of the $1/D$ expansion by taking the appropriate limit of 
the corresponding results for power-law potentials. For the two body case, 
this has already been done \ref\tom{U.~Sukhatme and T.~Imbo, Phys. Rev. D 
{\bf 28} (1983) 418\semi T.~Imbo, A.~Pagnamenta and U.~Sukhatme, Phys. Rev. D 
{\bf 29} (1984) 1669.}. The result is 
\eqn\five{E^{(D)}_{m_im_j}=(A/2)[\ell n(e\hbar^2D^2/4\mu Ar_0^2)+O(1/D)].}
The $D\to\infty$ limit is a {\it classical limit} (distinct from the $\hbar\to 
0$ limit), and the above result can be interpreted as $V_{{\rm eff}}(R_0)$ 
where $V_{{\rm eff}}$ is the large-$D$ effective potential. Here $R_0$ is the 
interquark distance at which this effective potential is minimized. In other
words, $R_0$ is the ``size'' of the meson. For the above log potential we have 
$R_0=\sqrt{\hbar^2/4\mu A}$.

In the three body case, the leading order large-$D$ result for power-law 
potentials is not completely analytic, as noted previously. However, the log
limit can still be recovered in closed form when two of the masses are equal. 
A careful analysis yields
\eqn\six{E^{(D)}_{mmm_3}=(3A/4)\{\ell n[e\hbar^2D^2(1+2x)(2a)^{1/3}/2mAr_0^2
(2-a)^2]+O(1/D)\} ,}
where $x=m/m_3$ and $a=8/[\sqrt{(2x+1)(2x+25)}+2x+5]$. This corresponds to a
minimum energy configuration of the large-$D$ effective potential in which the 
distance between $q_3$ and either of the other two quarks is ${\sqrt{\hbar^2
(1+2x)/2mA(2-a)^2}}$, and the angle $\theta$ between the two associated
displacement vectors satisfies ${\rm cos}(\theta )=1-a$. After some messy 
but straightforward algebra, one obtains the following expression for 
${\delta (mmm_3)=E^{(D)}_{mmm_3}-E^{(D)}_{mm_3}-E^{(D)}_{mm}/2}$:
\eqn\seven{\eqalign{\delta (mmm_3)=(A/4)[5&\ell n(\sqrt{(2x+1)(2x+25)}+2x+5)
+3\ell n(2x+1)\cr -6&\ell n(\sqrt{(2x+1)(2x+25)}+2x+1)-2\ell n(x+1)+O(1/D)].}}
Note that $\delta (mmm_3)$ is independent of $D$ as $D\to\infty$. When $x=1$, 
that is $m_3=m$, we have $\delta (mmm)=(3A/4)\ell n(4/3)\simeq (0.216)A$, which 
compares well with the numerical result of $(0.225)A$. (It is interesting to 
note that the large-$D$ result for $\delta (mmm)$ is {\it exactly} the same as 
that obtained in \ref\cohe{I.~Cohen and H.~J.~Lipkin, Phys. Lett. B {\bf 93} 
(1980) 56.}\ by seemingly independent means.) It seems reasonable to believe 
that at least some of this accuracy is due to a partial cancellation of the 
$1/D$ corrections between the two and three body cases. We can also easily 
extract the results for $x=0$ and $x\to\infty$ from \seven . We have $\delta 
(x=0)=(A/4)\ell n(3125/1458)\simeq (0.191)A$ and $\delta (x\to\infty )=(A/4)
\ell n(2)\simeq (0.173)A$. Scaling arguments imply that the latter result is 
actually exact! Unfortunately, in this limit we do not expect the log potential 
to provide an accurate description of quark dynamics. This is because we can 
think of the $x\to\infty$ limit as either $m$ getting very large or $m_3$ 
getting very small. In the former case, the two {\it very heavy} quarks of 
mass $m$ in the baryon are clearly in the perturbative ``Coulombic'' and not 
the logarithmic regime, while in the latter case a nonrelativistic treatment 
of the {\it very light} quark is not justified. (Similar problems do not occur 
in the $x\to 0$ limit viewed as $m_3$ getting very large.) As mentioned 
earlier, however, we {\it do} believe that the nonrelativistic log potential 
is applicable to systems containing $u,d,s,c$ and $b$ quarks. Here, $x$ can 
get as large as $m_b/m_n\simeq 15$.

It is often a bit more useful to consider the ratio $R(x)\equiv\delta (mmm_3)/
\delta (m_0m_0m_0)$. This quantity has the advantage of being independent of 
the parameters appearing in the log potential. A plot of $R(x)$ in the 
large-$D$ limit is shown in Fig.~1. It possesses a single critical point (a 
global maximum) in the range $0\leq x<\infty$. This occurs at $x=1$ where 
$R(1)=1$. (Note that the large-$D$ result for $R(x)$ is forced to be exact at 
$x=1$, so that one might expect it to be somewhat more accurate in the region
of physical interest than the corresponding result for $\delta$, which already 
seems to be within a few percent of the exact number for all $x$. More on this 
below. Of course, $R(x)$ is no longer exact as $x\to\infty$. But this is a 
trade-off that we will gladly accept since, as noted above, our model is not 
applicable at very large $x$.) The asymptotic values $R(0)\simeq 0.883$ and 
$R(\infty )\simeq 0.803$ represent {\it completely} parameter independent 
relations for the log potential, at least in the large-$D$ limit. Note also 
that $R(x)$ only changes by about 20\% over an infinite range of $x$ values, 
and only by about 10\% in the physical region $m_n/m_b\leq x\leq m_b/m_n$. 
Higher order $1/D$ corrections to the two and three body energies, and hence 
to $\delta$ and $R$, can in principle be computed. Although the resulting 
formulas will be more complicated, it is nice to know that there is a 
systematic analytic procedure for improving our results. For instance, using 
the results of \ref\gon{A.~Gonzalez, Few Body Sys. {\bf 10} (1991) 43\semi 
Y.~Y.~Goldschmidt, Nucl. Phys. B {\bf 393} (1993) 507.} one obtains 
\eqn\sevenp{\eqalign{\delta (mmm)&=(3A/4)[\ell n(4/3)+(2\sqrt{3}-\sqrt{2}-2)/D
+O(1/D^2)]\cr &\equiv [3\ell n(4/3)A/4]\{ 1+\alpha /D+O(1/D^2)\} .}}
For $D=3$ this yields $\delta (mmm)\simeq (0.228)A$ which is 99\% accurate, to 
be compared to the 96\% accuracy of the leading term alone. Of course, the 
above $1/D$ correction has no effect on $R(1)$ which is already exact at 
leading order. However, for $x\neq 1$ our result for $R(x)$ will have $1/D$ 
corrections. As an example, consider $R(0)$. A lengthy computation using the 
results of \mlod\ yields
\eqn\sevenpp{\eqalign{\delta (x=0)&=(A/4)[\ell n(3125/1458)+6(\sqrt{10}
+\sqrt{17}+2\sqrt{6}-5\sqrt{2}-5)/5D+O(1/D^2)]\cr &\equiv [\ell n(3125/1458)A
/4]\{ 1+\beta /D+O(1/D^2)\} .}}
For $D=3$ this gives $\delta (x=0)\simeq (0.202)A$, compared to the leading 
order result of approximately $(0.191)A$. Putting \sevenpp\ together with 
\sevenp\ we obtain
\eqn\sevenppp{R(0)=[\ell n(3125/1458)/3\ell n(4/3)]\{ 1+(\beta -\alpha )/D+
O(1/D^2)\} .}
For $D=3$ we have $(\beta -\alpha )/D\simeq 0.0016$, so that the change in
$R(0)$ due to $1/D$ corrections is about 1/6 of 1\%. ($R(0)$ changes from 
about 0.8833 to approximately 0.8848.) This is quite remarkable convergence. 
So although not exact as for $R(1)$, the leading order result for $R(0)$ 
should be extremely accurate. I believe the above examples to be illustrations 
of a quite general pattern: the leading order large-$D$ result for $\delta 
(mmm_3)$ is accurate to within a few percent, while the corresponding result 
for $R(x)$ is accurate to within a fraction of a percent --- at least over a 
large range of $x$. To be safe, one can compute the $1/D$ corrections to 
$R(x)$ for a general $x$ using the results of \ref\gont{A.~Gonzalez, Few Body 
Sys. {\bf 8} (1990) 73.}. However, it seems clear that the leading order 
results are accurate enough in general that the added complexity introduced by 
adding higher order corrections is not justified.

\epsfxsize=8cm
\centerline{\epsfbox{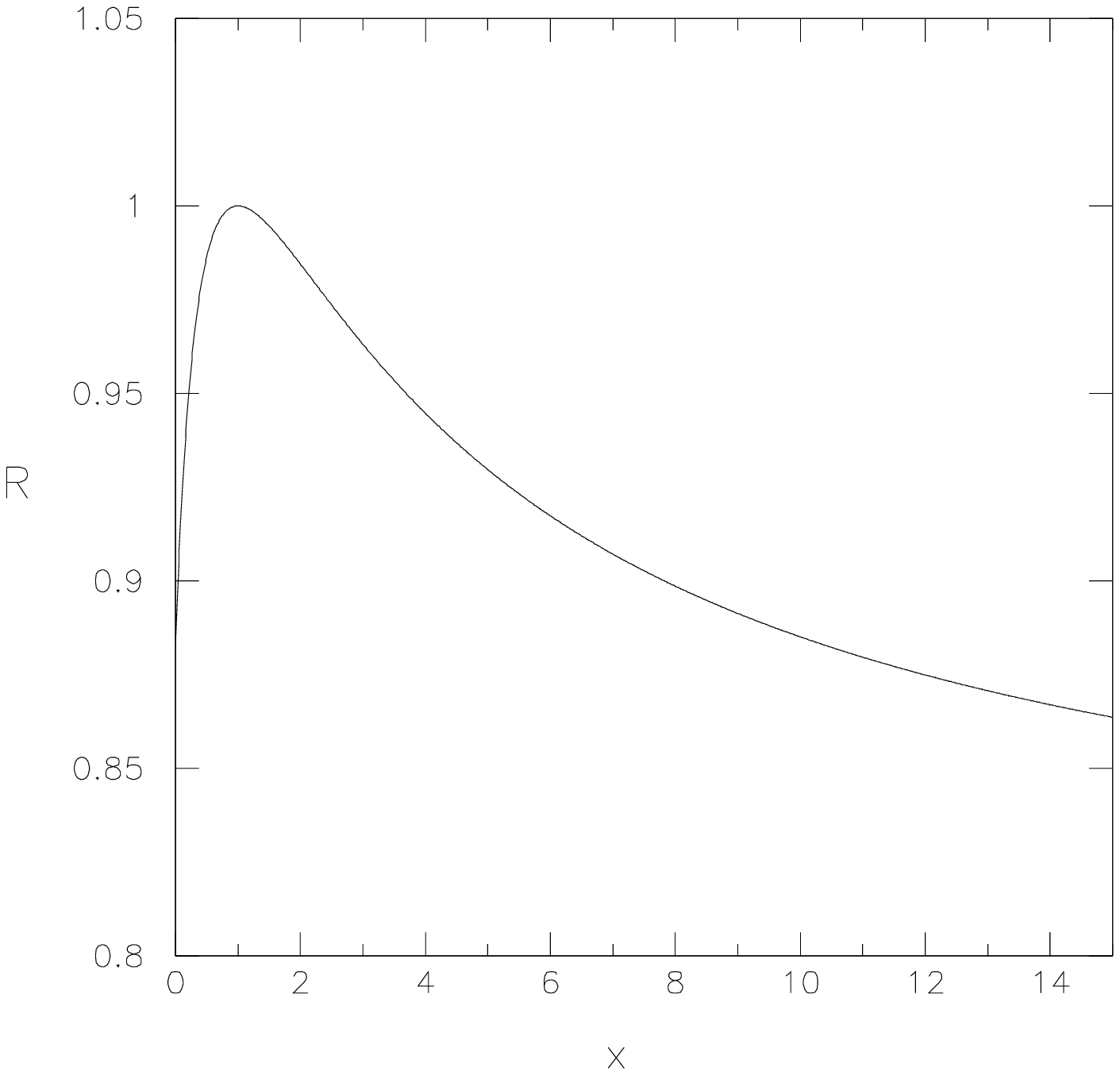}}
\vbox{%
{\centerline{Figure 1: $R(x)$ in the large-$D$ limit.}}}
\bigskip

How do the above results compare with observation? Lets first consider 
$x=m_n/m_s$ and $x=m_s/m_n$ (we have already discussed the situation 
for $x=1$). In the former case we have 
\eqn\eight{\eqalign{\delta (m_nm_nm_s)&=(M_{\Lambda}+M_{\Sigma})/4+
M_{\Sigma^*}/2 \cr &-(M_{\pi}+3M_{\rho})/8-(M_K+3M_{K^*})/4 \cr
&=171\pm 2\ {\rm MeV/c^2}.}}
Recalling that $\delta (m_nm_nm_n)=171\pm 2\ {\rm MeV/c^2}$ as well, this 
leads to the experimental value $R_{\rm exp}(m_n/m_s)=1.00\pm 0.02$. 
However, we cannot obtain an unambiguous theoretical prediction for $R(m_n/m_s)$ 
from \seven\ since $m_n$ and $m_s$ have not been determined in our model. One 
can use, among other things, the ratio of the $K$-$K^*$ and $\pi$-$\rho$ mass 
splittings as a naive estimate of $m_n/m_s$. This gives $m_n/m_s\simeq 0.63$. 
(For comparison, the ratio of the $\Sigma$-$\Sigma^*$ to the $N$-$\Delta$ 
splitting yields $m_n/m_s\simeq 0.65$.) But luckily, $R$ does not change too 
much within the entire range of ``reasonable'' values of $m_n/m_s$. For 
$0.5\leq m_n/m_s\leq 0.8$, $R(m_n/m_s)$ stays between 0.985 and 1.00, in 
good agreement with the experimental number. Using the corresponding range 
$1.25\leq m_s/m_n\leq 2.0$, the log model prediction for $R(m_s/m_n)$ also 
varies between 0.985 and 1.00. Unfortunately, we cannot obtain a purely 
experimental number to compare this with since certain states needed to find 
the required spin-averaged masses $M_{s{\bar s}}$ and $M_{ssn}$ do not exist. 
However, estimating the masses of these missing states from other 
spin-splittings of hadrons containing $u$, $d$ and $s$ quarks yields an 
``experimental'' value of $R(m_s/m_n)$ in the desired range.

We now turn to hadrons containing one heavy quark --- that is, one $c$ or $b$ 
quark. Taking the ratio of the $D$-$D^*$ and $\pi$-$\rho$ mass splittings 
yields the estimate $m_n/m_c\simeq 0.22$. (Using the recent measurement 
$M_{\Sigma^*_c}=2519\pm 2\ {\rm MeV/c^2}$ made by the CLEO collaboration 
\ref\cleo{CLEO Collaboration, preprint no. CLNS 96/1427 (1996).}, we can also 
estimate $m_n/m_c$ from the ratio of the $\Sigma_c-\Sigma^*_c$ and $N-\Delta$ 
splittings. This again gives $m_n/m_c\simeq 0.22$.) From \seven\ we then 
obtain the prediction $R\simeq 0.95$. As before, the deviation from this value 
is not too large over the entire range of reasonable values of $m_n/m_c$. More 
precisely, for $0.15\leq m_n/m_c\leq 0.3$, the value of $R$ varies from 
approximately 0.935 to 0.965. The experimental formula for $\delta (m_nm_nm_c)$ 
can be obtained from \eight\ by replacing $K$ and $K^*$ by $D$ and $D^*$, as 
well as substituting $\Lambda_c$, $\Sigma_c$ and $\Sigma_c^*$ for $\Lambda$, 
$\Sigma$ and $\Sigma^*$. This yields $\delta (m_nm_nm_c)=165\pm 2\ 
{\rm MeV/c^2}$, which corresponds to $R_{\rm exp}(m_n/m_c)=0.965\pm 0.02$, 
once more in good agreement with the theoretical prediction.

Similarly, the ratio of the $B$-$B^*$ and $\pi$-$\rho$ splittings yields 
$m_n/m_b\simeq 0.07$, which when plugged into \seven\ gives $R\simeq 0.91$. 
Moreover, for the range $0.05\leq m_n/m_b\leq 0.08$, the value of $R$ stays 
between 0.905 and 0.915. The experimental value can again be obtained from 
\eight , this time by substituting $B$ and $B^*$ for $K$ and $K^*$, and 
replacing $\Lambda$, $\Sigma$ and $\Sigma^*$ by $\Lambda_b$, $\Sigma_b$ and 
$\Sigma_b^*$. But since the experimental status of some of the bottom baryons 
is uncertain (see below), I instead use this formula in conjunction with the 
above result for $R$ to obtain a rather precise prediction for the $bnn$ center 
of gravity --- namely, $M_{bnn}=5774\pm 4\ {\rm MeV/c^2}$. To what extent can 
this result be checked? In other words, what is our current state of knowledge
about the $bnn$ baryons? As of a year ago, the $\Lambda_b$ had been seen but its
mass was not known very accurately, the world average being $5641\pm 50\ 
{\rm MeV/c^2}$. Since then, there have been three additional measurements. The 
most accurate is that of the CDF collaboration \ref\cdf{CDF Collaboration, Phys. 
Rev. D {\bf 55} (1997) 1142.}. They found a $\Lambda_b$ mass of $5621\pm 5\ 
{\rm MeV/c^2}$. The other two results are from the ALEPH \ref\alep{ALEPH 
Collaboration, Phys. Lett. {\bf B380} (1996) 442.} and DELPHI \ref\delp{DELPHI 
Collaboration, Phys. Lett. {\bf B374} (1996) 351.} collaborations. The ALEPH 
number is $5614\pm 21\ {\rm MeV/c^2}$, while DELPHI found $5668\pm 18\ 
{\rm MeV/c^2}$. Taking the weighted average of these new results and the previous 
world average gives a new world average for the $\Lambda_b$ mass of $5624\pm 5\ 
{\rm MeV/c^2}$, which we will use below. Recently, the DELPHI collaboration has 
also reported the first evidence for the $\Sigma_b$ and $\Sigma_b^*$ baryons 
\ref\delp{DELPHI Collaboration, preprint no. 95-107 PHYS 542 (1995).}. Their 
preliminary mass values\foot{I thank C.~Kreuter for informing me of the new 
(larger) error for $M_{\Sigma_b}-M_{\Lambda_b}$.} are $M_{\Sigma_b}-M_{\Lambda_b}
=173\pm 14\ {\rm MeV/c^2}$ and $M_{\Sigma_b^*}-M_{\Lambda_b}=229\pm 9\ 
{\rm MeV/c^2}$. Noting that $M_{bnn}=M_{\Lambda_b}+(M_{\Sigma_b^*}-
M_{\Lambda_b})/2 +(M_{\Sigma_b}-M_{\Lambda_b})/4$, we can plug in the above 
measurements and obtain the experimental value $M^{\rm exp}_{bnn}=5782\pm 8\ 
{\rm Mev/c^2}$. Though consistent with the above prediction, it would clearly be 
desirable to have a more precise measurement of all the $bnn$ baryon masses in 
order to make a better comparison. At this point it is also worth noting one 
strange feature of the measured values of the $\Sigma_b$ and $\Sigma_b^*$ masses 
--- namely, their difference.\foot{I thank Adam Falk for pointing this out to me, 
and Ben Grinstein for useful discussions.} We obtain from the DELPHI numbers that $M_{\Sigma_b^*}-M_{\Sigma_b}=56\pm 17\ {\rm MeV/c^2}$. An estimate of this 
difference from the spin-splitting pattern in the meson sector would be around 
$20\ {\rm MeV/c^2}$. (By contrast, recall the amazing consistency of the meson 
and baryon hyperfine splittings in the $snn$ and $cnn$ cases.) Various other 
simple estimates yield a similar number \ref\kwon{W.~Kwong and J.~L.~Rosner, 
Phys. Rev. D {\bf 44} (1991) 212.}. Given this discrepancy, I would not be 
surprised to see these preliminary mass values change somewhat in the near 
future and come into better agreement with both the mesonic spin-splittings 
and the predictions of the log model. 

It is worth stressing again that there is every reason to believe the above
predictions to be extremely trustworthy. As noted earlier, the nonrelativistic 
log model should provide an accurate description of the ground state hadrons 
for {\it some} values of the parameters, even for light quarks. However, all 
of the parameter dependence has cancelled out in $R(x)$ except for a 
relatively weak dependence on the quark mass ratio $x$, and we have incorporated 
our ignorance about these ratios into the overall uncertainty. Finally, $1/D$ 
corrections also seem to be completely under control --- the additional 
uncertainty in our results due to the neglect of such corrections is swamped 
by the quark mass uncertainties just mentioned. The appeal of the above 
approach is in its simplicity, accuracy and parameter-independence, and not 
in any rigorous connection to more fundamental ideas.

\vskip 12pt
\centerline{{\bf Acknowledgements}}
\vskip 12pt
It is a pleasure to thank Adam Falk, Ben Grinstein, Wai-Yee Keung, Young-Kee 
Kim, Christof Kreuter, Uday Sukhatme and William Wester for useful discussions.
This work was supported in part by the U.~S.~Department of Energy under 
contract number DE-FG02-91ER40676.

\listrefs
\bye